\documentstyle[11pt,moriond,epsfig]{article}

\def\be{\begin{equation}}
\def\ee{\end{equation}}
\def\bea{\begin{eqnarray}}
\def\eea{\end{eqnarray}}

\begin{document}
\title{TESTING $\Omega_0$ WITH X--RAY CLUSTERS: A PHYSICAL APPROACH}

\author{ PAOLO TOZZI }

\address{II Universit\`a di Roma \\
E-mail tozzi@roma2.infn.it}
\vspace*{0.6cm}

\maketitle\abstracts{
The X-ray emission from clusters of galaxies is one of the most pursued
observational probe to investigate the distribution of dark matter 
and the related density parameter
$\Omega_0$.  The crucial link to derive the statistics of observables 
from a dynamical theory is constituted by the physics for the diffuse 
baryons (or ICP) responsible of the X--ray emission.  Here we present a 
physical model for the ICP which leads to a definite $L$--$T$ relation.  
Then we perform a physically based cosmological test, pointing out three 
cold dark matter universes: a Tilted critical CDM, a flat CDM 
with $\Omega_0=0.3$, and an Open CDM with $\Omega_0=0.5$, which are discussed
on the basis of the RDCS survey.  }

\section{Introduction}

Groups and clusters of galaxies constitute cosmic structures sufficiently 
close to equilibrium and with sufficient density contrast ($\delta\approx 
2\, 10^2$ inside the virial radius $R$) as to yield definite 
observables.  They are dominated by dark matter (hereafter DM), while 
the baryon fraction is observed to be less than $20$\%.  The great majority 
of these baryons are in the form of {\sl diffuse plasma} (ICP) with densities 
$n\sim 10^{-3}$ cm$^{-3}$ and virial temperatures $k\,T\sim 
5$ keV, and are responsible for powerful X--ray luminosities $L\sim 10^{44}$ 
erg/s by optically thin thermal bremsstrahlung.  As the plasma is a good 
tracer of the potential wells, much better than member galaxies, the X--ray 
emission is a powerful tool to investigate the mass distribution out to 
moderate and high redshifts.  The ICP temperature directly probes the height 
of the potential well, with the baryons in the  role of  mere tracers; 
on the other hand, the luminosity with its strong dependence on density  
($L\propto n^2$) reliably probes the baryonic content and distribution.  
Statistically, an average $L$--$T$ correlation is observed along with
substantial scatter, and this provides the crucial link to relate the X-ray 
luminosity functions with the underlying statistics of the DM.  

A physical model for the diffuse baryons is difficult to achieve.  
In fact, the simple self similar model (Kaiser 1986), which assumes the ICP 
amount to be proportional to the DM's at all $z$ and $M$, leads to a relation 
$L\propto T^2$, conflicting with the observed correlation for rich clusters.  
The latter is close to $L\propto T^{3.5}$ (David et al. 1993; 
Mushotzky \& Scharf 1997).  Here we propose a physical model for baryons, 
which leads to a prediction for the $L$--$T$ relation (see Cavaliere, Menci
\& Tozzi 1997, CMT97) and allows a non parametrical approach
to the search for cosmological parameters.  The results, presented in \S 3, 
are a synthesis from Cavaliere, Menci \& Tozzi (1998, CMT98).  

\section{A physical model for the ICP}

Respect to the self--similar model, we indicate the 
missing ingredient in the stellar energy feedback by supernovae.  We assume 
that such non gravitational heating is efficient in depleting the
potential wells of the clusters progenitors, at $z\simeq 1 \div 2$, and in
pre--heating the intergalactic medium to temperatures in the range $T_1=0.1
\div 0.8$ keV, as recently observed in the outer cluster atmosphere
(Henriksen \& White 1996).  

We describe clusters evolution as a sequence of hierarchical merging episodes 
of the DM halos; the history of such episodes is followed in the framework 
of the hierarchical clustering by Monte Carlo simulations based on merging 
trees consistent with the Press \& Schechter (1974) statistics.  
In the ICP, shocks of various strengths (depending on 
the temperature ratio between the accreted gas and
the plasma in the main progenitor) are associated to such merging events.  
The shocks provide the boundary conditions for the ICP to 
re--adjust to a new hydrostatic equilibrium.  

\subsection{Hydrodynamical equilibrium}

In the framework of the hierarchical clustering it is possible to check the 
assumption of equilibrium.  Roettiger, Stone \& Mushotzky (1997) showed 
that after a major merging event, i.e., 
with a mass ratio less than $2.5$, the non thermal contribution to the pressure 
is negligible after two Gyrs from the epoch of merging.  
Adopting this as a conservative rule to identify {\sl disturbed } clusters 
at redshift $z=0$, we find that the fraction of such clusters 
(for which the hydrodynamical equilibrium does not fully apply) is always less 
than $20$\% in most CDM universes.  

The next step is computing the disposition of the ICP in 
equilibrium in a DM potential well.  Then we simply use the hydrostatic 
equilibrium equation $ ({{dP}/ {dr}})/n=-G {{M(<r)}/{r^2}}$, 
where $n(r)$ is the gas density profile, $M(<r)$ is the mass contained within 
the radius $r$ (dominated by DM) and $P$ is the pressure.  
To solve the equation we need the boundary condition, i.e., the density
of the gas at the virial radius $n(R)$, and the state equation, that we
write for a polytropic gas as $P\propto T\, n^\gamma$, where
$\gamma$ is the polytropic index treated as a free parameter.  
The resulting profile is: 
\begin{equation}
n(r)=n(R)\Big[ 1+ \beta \Big( {{\gamma -1}\over{\gamma}}\Big)
\Big( \phi(R)-\phi(r)\Big)\Big]^{1/(\gamma -1)}\, ,
\label{prof}
\end{equation}
where $\phi\equiv V/\sigma_r$ is the adimensional gravitational potential, 
and $\beta\equiv \mu m_H \sigma_r/kT_2$, with $\sigma_r$ the line--of--sight
velocity dispersion of the DM particles.  Here $T_2$ is ICP the temperature 
inside the virial radius.  The above equation can be considered a generalized 
$\beta$--model (Cavaliere \& Fusco--Femiano 1978), which reconduces 
to the usual isothermal case for $\gamma\rightarrow 1 $.  Thus, before 
computing the ICP distribution, we need
the boundary condition $n(R)$.  

\subsection{Physics of shocks}

\begin{figure}
\centerline{\psfig{figure=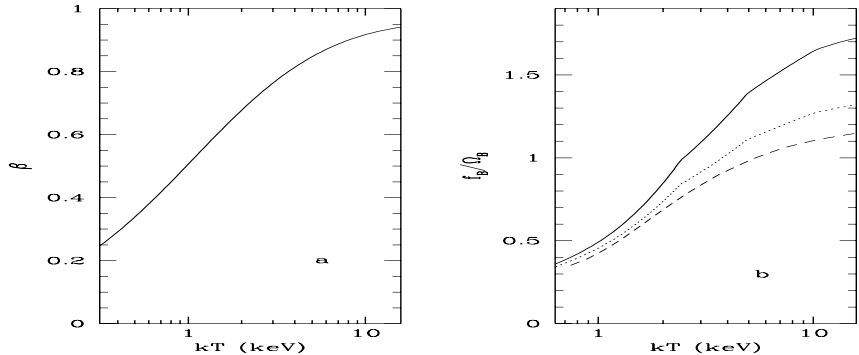,height=5.truecm,width=12truecm}}
\caption{\sl a) The $\beta(T)$ parameter entering equation \ref{prof}.  
b) Baryonic fraction respect to the universal value $\Omega_B$ for
different values of $\gamma$.  Continuous line: $\gamma =1$; dotted line:
$\gamma =1.1$; dashed line: $\gamma = 1.2$.  }
\label{fig1}
\end{figure}

We expect the inflowing gas (with velocity $v_1$) to become 
supersonic close to $R$. In fact, many hydrodynamical simulations (see, e.g., 
Takizawa \& Mineshige 1997) show 
shocks to form, to convert most of the bulk energy into thermal energy, 
and to expand slowly remaining close to the virial radius.  So 
we take $R$ as the shock position, and focus on nearly static 
conditions inside, with the internal bulk velocity $v_2<< v_1 $.  

The post-shock state is set by conservations across the shock 
not only of the energy, but also of mass  and momentum, as described by the  
Rankine-Hugoniot conditions (see Landau \& Lifshitz 1959).  
These provide at the boundary the temperature jump $T_2/T_1$, and the
corresponding density jump $g\equiv n(R)/n_1$, which  reads:  
\begin{equation}
g\Big({T_2\over T_1}\Big) = 
2\,\Big(1-{T_1\over T_2}\Big)+\Big[4\, 
\Big(1-{T_1\over T_2}\Big)^2 + {T_1\over T_2}\Big]^{1/2}
\label{gt}
\end{equation}
for a plasma with three degrees of freedom.  
Here $n_1$ is the baryon density external to the virial radius, and it
is assumed to be unbiased respect to the universal value, i.e., $n_1=\Omega_B 
\rho$ where $\rho$ is the total density.  Eq. \ref{gt} includes both {\it weak} 
shocks (with $T_2 \approx T_1$, appropriate 
for small groups accreting preheated gas, 
or for rich clusters accreting comparable clumps), 
and {\it strong} shocks (appropriate to ``cold inflow" as in rich 
clusters accreting small clumps and diffuse gas).  
Given the nearly static post-shock condition $v_2<< v_1$, it is possible to 
show that in the case of strong shocks 
$k\,T_2\approx -V(R)/3+3k\,T_1/2$ holds, where the second term is
the contribution from non gravitational energy input (CMT98).  
This leads to a factor $\beta(T)$ which decline from $\sim 1$ for rich 
clusters, to $\sim 0.4$ for poor groups, where the nuclear competes with the 
gravitational energy (see fig. \ref{fig1}a).  A specific gas profile depends 
on the choiche for the potential well (in the following we use forms given
by Navarro, Frenk \& White (1996), but 
this is not mandatory).  A general result is that the corresponding baryonic 
fraction lowers down with the mass scale by a factor of three from clusters 
to groups (see fig. \ref{fig1}b).  

\subsection{The $L$-$T$ correlation}

\begin{figure}
\centerline{\psfig{figure=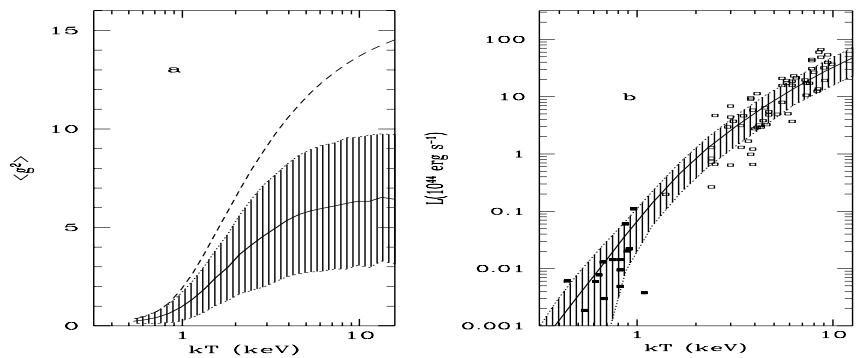,height=5truecm,width=12truecm}}
\caption{\sl a) The factor $\langle g^2 \rangle$ averaged over merging histories
(solid line) with $96$\% confidence level (shaded region).  The non--averaged
$g$ is plotted for comparison (dashed line). b)  Predicted
$L$--$T$ relation.  Data by David et al. (1993)
for clusters (open squares) and Ponman et al. (1996) for 
groups (solid squares). }
\label{lt}
\end{figure}

The X-ray luminosity of a cluster with temperature profile $T(r)$ and density 
profile $n(r)$ can be written: 
\begin{equation}
L\propto \langle g^2(T) \rangle \int d^3 r\,{{n^2(r)}\over {n^2(R)}}\, 
T^{1/2}(r)\, \, , 
\end{equation}
where the statistical effect of the merging histories has to be taken 
into account.  In fact, for a cluster or a group of a given mass (or 
temperature), the effective compression factor squared  $\langle g^2 \rangle$ 
is obtained upon averaging eq. \ref{gt} over the sequence of the DM 
merging events; in such events, $T_2$ is the virial temperature of 
the receiving structure, and $T_1$ is the higher  between the stellar 
preheating temperature and the virial value prevailing in the clump being 
accreted.  The averaged $\langle g^2 \rangle$ is lower than the 
$g^2$ computed with a single temperature $T_1$, 
because in many events the accreted gas has a temperature higher than the 
preheating value.  In addition, an intrinsic {\it variance} 
is generated from the merging histories (see fig. \ref{lt}a).  

In agreement with the observations, the {\it shape} of the average $L-T$ 
relation flattens from $L\propto T^5$ at 
the group scale (where the nuclear energy from stellar preheating 
competes with the gravitational energy) to $L\propto T^3$ at the 
rich cluster scales (see fig. \ref{lt}b). At larger temperatures the shape 
asymptotes to 
$L\propto T^2$, the self-similar scaling of pure gravity.  Notice the 
intrinsic {\it scatter} due to the variance in the dynamical merging histories, 
but amplified by the $n^2$ dependence of $L$.  The average normalization 
rises like $\rho^{1/2}(z)$, where $\rho$
is the effective external mass density which increases as $(1+z)^2$ 
in filamentary large scale structures hosting most groups and clusters
(see Cavaliere \& Menci 1997).  We check that 
the shape of the $L$--$T$ relation 
is little affected by changes of $\gamma$ (we assume a fiducial
value $\gamma =1.2$, see CMT98).  

\section{A physically based cosmological test}

\begin{figure}
\centerline{\psfig{figure=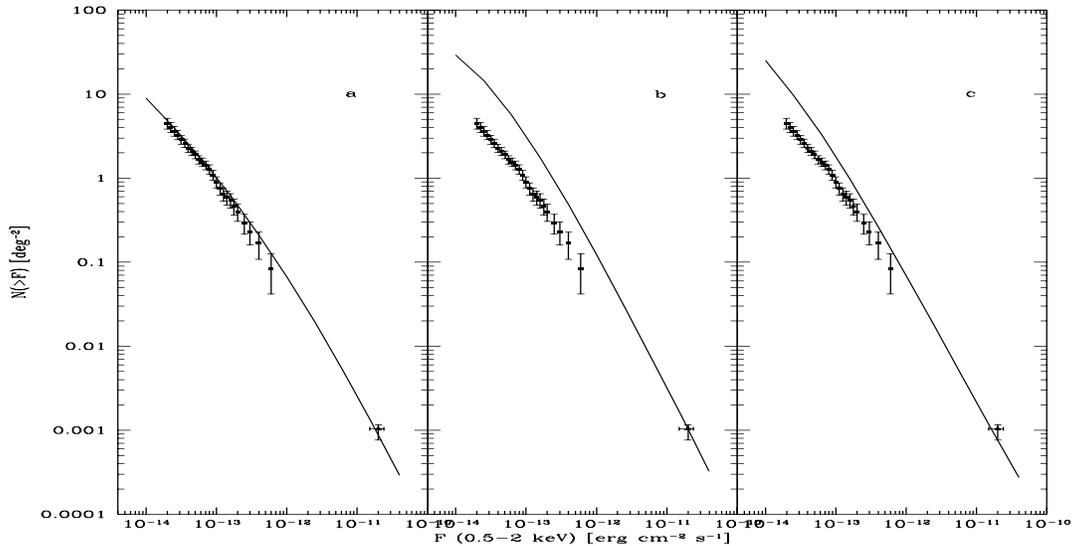,height=8truecm,width=15truecm}}
\caption{\sl Predicted counts for TCDM (a), OCDM (b) and $\Lambda$CDM (c).  
Data by Rosati et al. (1998), and Piccinotti et al. (1982) for the point
at high fluxes.  }
\label{counts}
\end{figure}

We adopt the Press \& Schechter rendition of the hierarchical clustering.  
For each CDM universe we first check agreement with local constraints on 
the luminosity function $N(L)$ and on the temperature function $N(T)$; 
then the test is performed on the flux counts $N(>F)$ in the ROSAT band, 
which include the effect of the evolution (the higher $z$ sample of RDCS 
by Rosati et al. 1998).  While the normalization $\sigma_8$ is fixed by 
the constraints from COBE for a given CDM spectrum, the predictions for the 
faint counts are sensitive to $\Omega_0$.  

We focus on three popular CDM universes.  The first is
the critical TCDM, for which we adopt the tilted spectrum with primordial 
index $n_p=0.8$, with amplitude $\sigma_8=0.66\,(1\pm 0.08)$, and with a 
high baryonic fraction $\Omega_B=0.15$ (the Hubble constant is $h=0.5$).  
The tilt is chosen so
as to minimize one of the main problems of the Standard CDM, namely, the 
excess of small-scale power.  The high baryon fraction is chosen to solve
the so called ``baryonic crisis'' (White et al. 1993).  
We note that the slope of predicted 
counts is sufficiently {\it flat} to fit both
the bright and the faint data by lowering $\sigma_8$ to within the COBE 
uncertainty (in fig. \ref{counts}a we adopt $\sigma_8=0.6$).  

\begin{figure}
\centerline{\psfig{figure=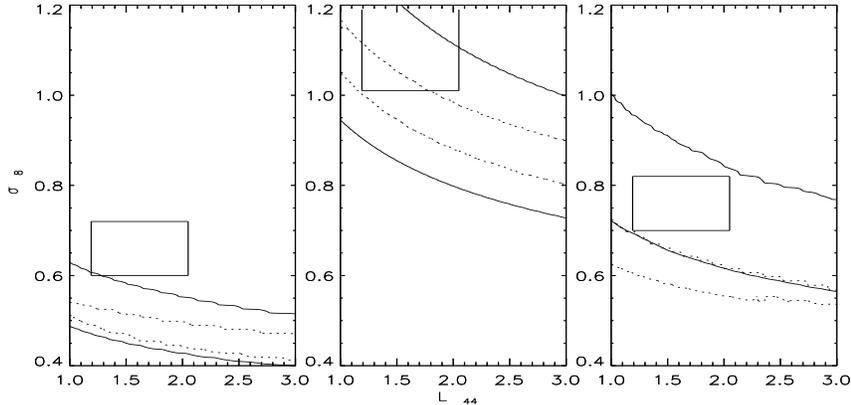,height=6truecm,width=12truecm}}
\caption{\sl The $99$\% confidence contours for both the computed 
local luminosity function (solid lines) and the computed number counts 
(dotted lines), in the $L_{44}-\sigma_8$ plane ($L_{44}=L_o/10^{44}$ erg/s). 
The boxes indicate $\pm 1$ standard deviations in $\sigma_8$ (corresponding 
to the COBE uncertainty) and in $L_{44}$.  a) TCDM, consistent with the counts 
within  $2$ standard deviations below $\sigma_8$; b) $\Lambda$CDM; c) 
OCDM ($\Omega_o=0.5$). 
}
\label{ell}
\end{figure}

The second is an open CDM universe.  After Liddle et al. (1996), we focus first 
on the representative OCDM cosmogony, with
$\Omega_o=0.5$, $h=0.65$ with $\Omega_B=0.07$, which yield 
$\sigma_8=0.76\,(1\pm 0.08)$.  It is seen that the 
counts show excesses over the data.  
The underlying reason is that in open cosmologies long lines of sight and slow
dynamical evolution conspire to yield a slope of the counts 
too {\it steep} to account for both faint and bright counts.  

The third universe ($\Lambda$CDM) has a flat geometry with $\Omega_o=0.3$ and 
$\Omega_{\lambda}=0.7$, $\Omega_B=0.05$ and $h=0.7$.  The normalization 
is $\sigma_8=1.1\,(1\pm 0.08)$.  In this intermediate condition
faint counts are higher respect to TCDM, yet lower than in the 
$\Omega_o\approx 0.5$ case, to yield a moderate excess (fig. \ref{counts}c).  

For a synthetic presentation, we also show in fig. \ref{ell} the effects of 
varying $\sigma_8$ and the normalization of the $L-T$ relation, i.e., the
average luminosity $L_{44}$ corresponding to $4.5$ keV. In TCDM and in 
$\Lambda$CDM the counts are consistent with 
the observations on considering the uncertainties in the present COBE 
normalization and the intrinsic uncertainty in the $L$-$T$ relation, while, 
on the other hand, in OCDM the counts are inconsistent with local data
by a significant excess.  

\section{Discussion and conclusion}

We presented a physically based approach to cosmological tests with clusters 
of galaxies.  We describe the X--ray emission from clusters with a specific 
model for the diffuse baryons.  In this sense this approach is alternative 
to the parametrical approach (see Borgani this meeting).  

The results of our model depend on two parameters, the external temperature 
$T_1$ and density $n_1$, which are {\sl not} free. Specifically, 
we use for $T_1$ the range $0.1\div 0.8 $ keV provided by the stellar 
preheating.  The value of $n_1$ for rich clusters is related to the DM 
density by the universal baryonic fraction.  Thus we compute the 
expression of the bolometric luminosity for a given temperature.  The 
average of the square of the density jump factor $\langle g^2 \rangle$ 
over the merging histories coupled
with $\beta(T)$ is what gives to the statistical $L-T$ correlation the 
curved shape shown in fig. \ref{lt}b. 
In addition, our approach predicts an intrinsic {\it variance} of dynamical
origin due to the different merging histories, and built in the factor $g^2$.  

With the ICP state so described, we proceeded to constrain 
the cosmological parameters. After the observations 
by Rosati et al. (1998), we have computed the X-ray observables for 
groups and clusters of galaxies.  On the basis of local data, 
the set of acceptable CDM universes is restricted to three disjoint 
domains: 
$\Omega=1$ for the Tilted CDM with high baryon content; 
$\Omega_o\simeq 0.5$ for standard CDM; $\Omega_o\approx 0.3$ for CDM in 
flat geometry.  However, only the TCDM and the $\Lambda$CDM universes
give acceptable faint counts.  
As an overall remark, a common feature of all the above 
universes is constituted by some excess in the counts.  This may 
indicate some non--trivial incompleteness in the canonical hierarchical 
clustering, worth keeping under scrutiny.  We recall that in the adiabatic 
models for the ICP (Evrard \& Henry 1991, 
Kaiser 1991) the evolution of the $L-T$ relation is reduced or even
negative, thus alleviating the excess.  However, the anti--evolution 
required in OCDM would be very difficult to justify (the adiabatic models
are largely discussed in CMT98).  

Now the question is: to what extent enlarging the data base on X--ray 
clusters will help in further constraining cosmology?  We argue that 
the variance intrinsic to the hierarchical clustering, and amplified by the ICP 
emissivity, sets an effective limitation.  Richer, faint surveys 
will hardly provide a sharper insight into cosmology unless one reduces both 
the uncertainty concerning $\sigma_8$ and the larger one concerning $L_{o}$.  
However, we stress that such efforts will find soon a more proper aim 
than constraining $\Omega_o$.  This is because MAP, and subsequently PLANCK, 
will accurately measure on still linear scales not only the perturbation 
power spectrum but also directly $\Omega_o$.  Once the cosmological framework 
has been fixed, the study of groups and clusters in X-rays 
will resume its proper course, that is, the physics of systems of intermediate 
complexity which is comprised of the DM and of the ICP component.

\section*{Acknowledgments}
The author acknowledge partial founding by the Training Mobility Research 
Programme.

\end{document}